\documentclass[pre,floatfix,twocolumn, 10pt]{revtex4-2}
\usepackage{graphicx}
\usepackage{amssymb}
\usepackage{natbib}
\usepackage{dcolumn}
\usepackage{bm}
\usepackage[normalem]{ulem}
\usepackage{color}
\usepackage[colorlinks=true, linkcolor=blue, citecolor=blue]{hyperref}
\usepackage{subfigure}
\usepackage{graphicx,psfrag,xspace}
\usepackage{color}
\usepackage{hyperref}
\usepackage{amssymb,amsfonts,amsmath}
\usepackage{setspace}

\begin{document}

\title{Nonholonomic constraints at finite temperature}

\author{Eduardo A. Jagla}
\affiliation{Centro At\'omico Bariloche, Instituto Balseiro,
Comisi\'on Nacional de Energ\'ia At\'omica, CNEA,
CONICET, UNCUYO, Av.~Bustillo 9500 (R8402AGP) Bariloche, R\'io Negro, Argentina}

\author{Anthony M. Bloch}
\affiliation{Department of Mathematics,
University of Michigan, Ann Arbor, MI 48109, USA}

\author{Alberto G. Rojo}
\affiliation{Department of Physics, Oakland University, Rochester, MI 48309, USA}

\date{today}

\begin{abstract}
We investigate the behavior of dynamical systems with nonholonomic constraints when coupled to a thermal
bath, focusing on the paradigmatic case of the Chaplygin sleigh. A straightforward Langevin-type approach---obtained by naively adding stochastic and dissipative terms to the equations of motion---predicts a regime in
which useful work can be extracted, violating the second law of thermodynamics. To resolve this paradox, we
resort to a physically motivated implementation of the nonholonomic constraint as the limiting case of a viscous
interaction. However, at finite temperature, fluctuation-dissipation relations imply that the viscous force has to be
complemented with stochastic forces acting at the contact. We show that their incorporation restores compliance
with the second law. Therefore, our results place fundamental limits on the physical realizability of idealized
nonholonomic constraints.
\end{abstract}
\maketitle

\section{Introduction}

Constraints on mechanical systems are classified as holonomic or nonholonomic.
Holonomic constraints are constraints on the configuration space, which ultimately reflect a redundancy in the number of
degrees of freedom used to define the state of the system.
On the other hand, nonholonomic constraints (NHCs)
are coordinate dependent restrictions on the velocities, and are non-integrable in the sense
that they cannot be rewritten as constraints on the coordinates alone.

A typical example of a NHC is the contact of a sphere or a knife
edge with a plane. If the plane is at rest, then the velocity of the contact point (for the case of the sphere), or the component of the velocity perpendicular to the edge (in the knife edge case) must vanish. The equations of motion in the presence of NHCs are given by the Lagrange-D'Alembert principle.
A remarkable feature of certain nonholonomic systems is that, while their dynamics conserve energy, they are nevertheless dissipative or irreversible in the sense that they do not conserve phase-space volume \cite{Bloch2003,NeFu1972,ZeBl2003,ArKoNe1988,BlMaZe2005,BlMaZe2009a,BlKrMaMu1996}.

There is extensive research on the dynamics of nonholonomic systems (see \cite{Bloch2003,NeFu1972,ArKoNe1988} and references therein). Here we explore
the behavior of systems with NHCs when they are in contact with a thermal bath.
Related references on nonholonomic mechanics and diffusion are \cite{HoRa2015,stoch_chap},
and the sleigh interacting with its environment in a non-stochastic setting
is discussed, e.g., in \cite{FeTa2018,FeGa2010,SaZo2020}, and a stochastic analysis of nonholonomic systems related to our work is \cite{GaYo2023}. Ref.~\cite{stoch_chap} discusses the dynamics of the stochastic Chapylgin sleigh.
Our analysis, on the other hand, focuses on the relationship
to the second law of thermodynamics. Other interesting related work includes the Gaussian thermostat discussed in, e.g., \cite{DeMo1996,RoBlo2009}
where there is a nonlinear nonholonomic constraint.

In this paper we consider
the dynamics of a paradigmatic system with a NHC
in equilibrium with a thermal bath.
Choosing an appropriate anisotropic (yet consistently realizable) interaction to the bath,
we first show that if the constraint is taken at face value the system violates the second law of thermodynamics.
We provide a detailed explanation of the mechanism behind this violation and subsequently proceed to
solve the enigma.
Our main conclusion is that when physically implementing a NHC, to impose the dynamical constraints on the equations of motion
may not suffice.

Formally, the key point is noticing that in parallel with the mathematical difference between holonomic and nonholonomic constraints, there are also important differences in the
way each kind of constraint can be physically implemented. Holonomic constraints, given in general as $V(x_1, ..., x_n)=0$ (where $x_i$ are the coordinates of the degrees of freedom of the system) can be implemented by the inclusion of potential energy terms in the Lagrangian of the form $kV^2/2$, and taking the limit $k\to\infty$. This corresponds physically to adding the
energy of a ``spring'' with stiffness $k$. As $k\to \infty$ we recover the constraint $V=0$.
For NHCs this technique does not work, as the constraints are nonintegrable.
One way of physically implementing a NHC
is to consider it as a limiting case of a friction force with a friction coefficient $\mu$ \cite{Caratheodory1933,Fufaev1964}. The NHC is obtained when $\mu\to\infty$. This naive implementation corresponds to the zero-temperature limit. At finite temperature,
however, it becomes necessary to account for the fact that the frictional force is accompanied by thermal fluctuations, which must comply with the fluctuation-dissipation \cite{Kubo} relation.
Thus, when the system is coupled to a thermal bath, a stochastic Langevin force must be retained in the constraint; there is an apparent violation of the second law of thermodynamics. The main result of our paper is to show that the inclusion of this term restores the thermodynamic consistency of the model.

Although our analysis applies to a variety of nonholonomic systems, for clarity of presentation we focus on the case of the Chaplygin sleigh \cite{Chaplygin1949}---a prototypical nonholonomic system---when coupled to a thermal bath. In Sec.~\ref{EPSsection} we briefly discuss
more general Euler--Poincar\'e--Suslov systems \cite{Bloch2003,Kozlov2015} and in particular the Suslov problem, which has similar
dynamics.

\section{Model and Results}

\subsection{The Chaplygin sleigh}

\begin{figure}[h]
\includegraphics[width=8cm,clip=true]{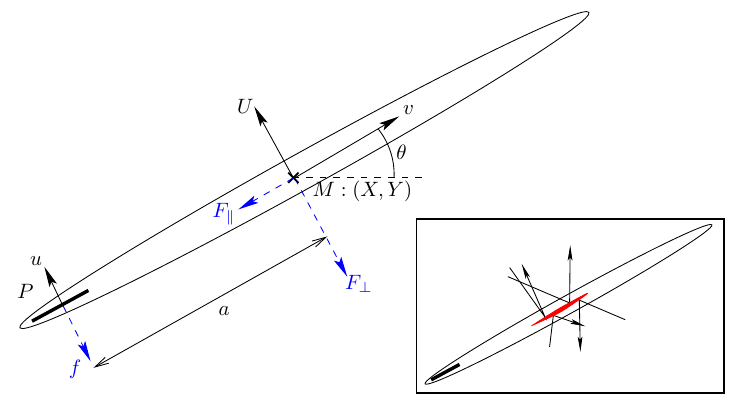}
\caption{Geometry of the Chaplygin sleigh. The center of mass is at point $M:(X,Y)$. The standard NHC at $P$ requires
the velocity $u$ perpendicular to $PM$ to be zero. The inset shows the way in which a ``sail'' (in red) is included, to exchange energy with the particles of a thermal bath.
}\label{f1}
\end{figure}

Here we present a classic example of a system with a NHC, namely, the Chaplygin sleigh \cite{Bloch2003,Chaplygin1949}.
It consists (see Fig.~\ref{f1}) of a rigid body moving on a two-dimensional surface. This motion is frictionless everywhere, except
at a single pivot point $P$ (that does not coincide with the center of mass M). The
movement of $P$ is assumed to occur along the $PM$ line only.
Physically, this constraint can be thought to be implemented by a skate blade located at P.
We introduce a force $f$ at P, along the direction perpendicular to $PM$ that will be used to enforce the constraint.
We note that $m$ and $I$ are the mass and moment of inertia of the body, $a$ is the distance between $P$ and M, and $\theta$ the angle between $PM$ and a fixed spatial direction $\hat x$.
The equations of motion of the body in the absence of external forces ($F_\perp=F_\parallel=0$) can then be readily written in Newtonian form as
\begin{subequations}
\begin{align}
m\ddot X&=f\sin(\theta), \label{xy_a}\\
m\ddot Y&=-f\cos(\theta) \label{xy_b}
\end{align}
\label{xy}
\end{subequations}
for the displacement of the center of mass, and the torque equation
\begin{equation}
I\ddot \theta= af.
\label{torque}
\end{equation}

\begin{figure}[ht]
\includegraphics[width=7cm,clip=true]{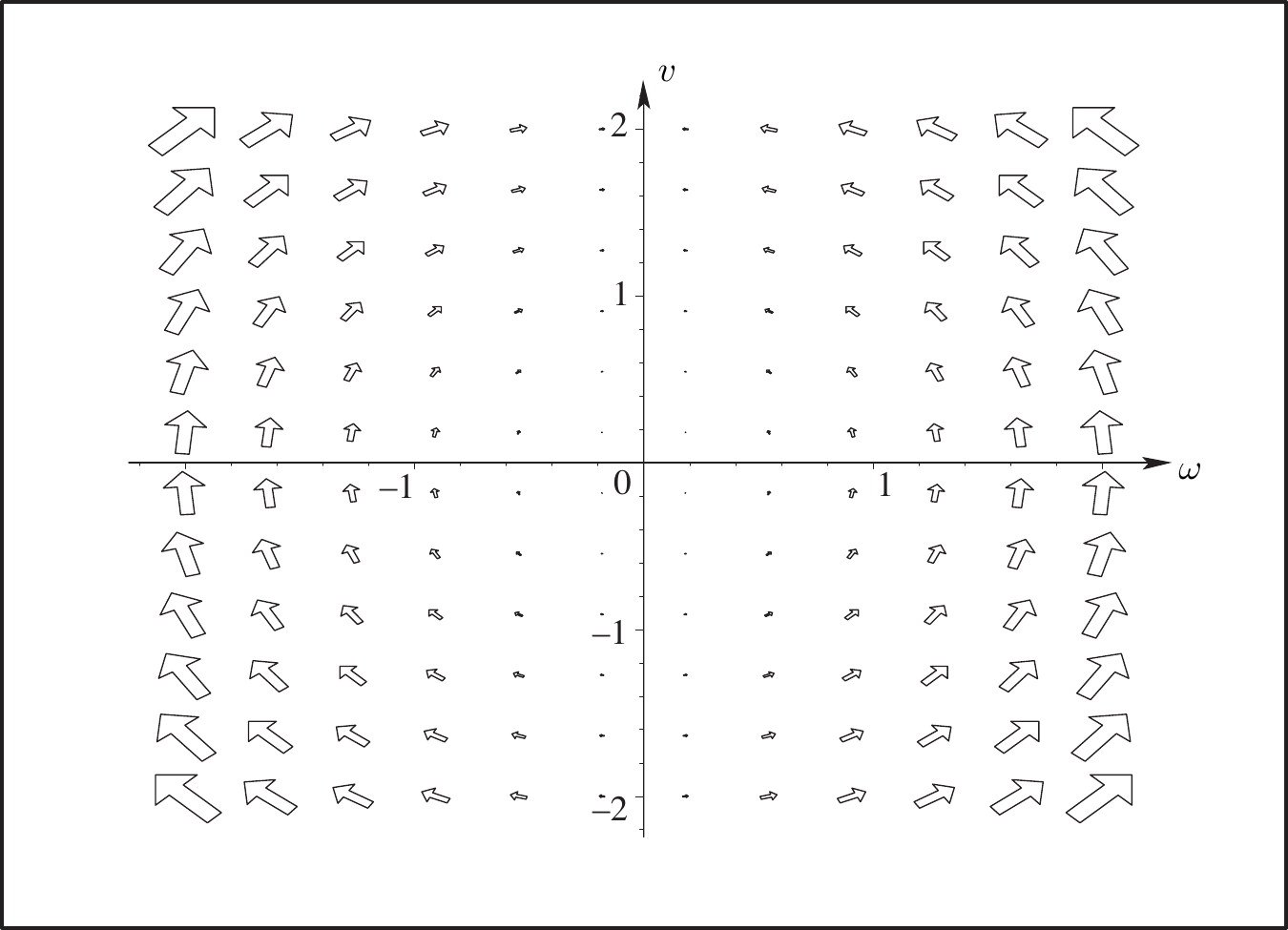}
\caption{Phase portrait of the Chaplygin sleigh.
}\label{Phase}
\end{figure}

The components of the velocity of the center of mass along, and perpendicular to $PM$ will be noted as $v$ and $U$, respectively. The
NHC implemented at $P$, implies that the velocity of point $P$ perpendicularly to $PM$
is zero. The Chaplygin sleigh equations are obtained when $f$ is taken to be the force
needed to impose this constraint.
We obtain in this case (see Appendix~\ref{AppA})
\begin{subequations}
\begin{align}
\dot v&=a\omega^2, \label{clasico_a}\\
\dot\omega&=-\frac{v\omega}{ \frac I{am}+a}
\end{align}
\label{clasico}
\end{subequations}
($\omega\equiv \dot \theta$).

These equations have the property \cite{Bloch2003} that any initial condition $(v_0,\omega_0\ne0)$ will converge
as $t\to\infty$ to the state
$(v_\infty,0)$, with $v_\infty=\sqrt {v_0^2+( a^2+I/m)\omega_0^2}$, namely, a state of no rotation, in which the initial rotational energy has been converted to translational kinetic energy, as shown in the phase portrait of Fig.~\ref{Phase} \cite{Bloch2003}. Note that the asymptotic state
is reached from an infinite set of possible initial conditions: the phase space volume $d\Omega\equiv dv\, d\omega$ is not conserved during evolution rendering the system ``dissipative'' (and non-measure-preserving).

\subsection{Coupling to a thermal bath}

A Chaplygin sleigh coupled to a thermal bath at temperature $T$ has been considered in \cite{stoch_chap}.
The procedure is to add to the Chaplygin equations of the previous section stochastic forces $F_\parallel$ and $F_\perp$ (see Fig.~\ref{f1}) along, and perpendicular to PM, representing the coupling to a thermal bath.
The Newton equations in this case read
\begin{subequations}
\begin{align}
m\ddot X&=(f+F_\perp)\sin(\theta)-F_\parallel\cos(\theta), \label{nuevas_a}\\
m\ddot Y&=-(f+F_\perp)\cos(\theta)-F_\parallel\sin(\theta), \label{nuevas_b}\\
I\ddot \theta&= af. \label{nuevas_c}
\end{align}
\label{nuevas}
\end{subequations}
Notice that we assume that the coupling to the thermal bath occurs only in a region very near the center of mass, and therefore
$F_\parallel$ and $F_\perp$ have a negligible effect on the torque balance.
From here we can derive the equations of motion in the form (see Appendix~\ref{AppA})
\begin{subequations}
\begin{align}
\dot v&=a\omega^2-\frac{F_\parallel}m, \label{2l_a}\\
\dot\omega&=-\frac{am}{I+a^2m}  \left ({v\omega+\frac {F_\perp}m}\right ).
\end{align}
\label{2l}
\end{subequations}
$F_\parallel$ and $F_\perp$ contain stochastic and deterministic parts that are related through a
fluctuation-dissipation relation. Concretely,
\begin{subequations}
\begin{align}
F_\parallel&=-\lambda_\parallel v +\sqrt{2\lambda_\parallel k_BT }\,\eta_\parallel(t), \label{2f_a}\\
F_\perp&=-\lambda_\perp \omega a +\sqrt{2\lambda_\perp k_BT }\,\eta_\perp(t),
\end{align}
\label{2f}
\end{subequations}
where $\eta_\perp(t)$ and $\eta_\parallel(t)$ are white noise terms, with zero mean and unitary variance.
An isotropic coupling to the bath (as used in \cite{stoch_chap}) requires $\lambda_\parallel=\lambda_\perp$.
However, here we will allow for the two coupling constants to be different. Jung {\em et al.} obtain that with
$\lambda_\parallel=\lambda_\perp$ the mean value of $v$ is different from
zero as long as $a\ne 0$, namely, there is a tendency of the sleigh to move in the PM direction. Yet they point out that
the system is {\em thermalized} in such a way that the average total kinetic energy of the system is $k_BT$. However, we find that the situation is different if $\lambda_\parallel\ne\lambda_\perp$. In fact, an apparent violation of the second law of thermodynamics can occur, particularly if $\lambda_\parallel\ll \lambda_\perp$.
We emphasize that it is perfectly possible to physically implement a situation with different coupling, parallel and perpendicular to the sleigh. In fact, we can think of adding a ``sail'' of negligible mass, located at the center of mass of the body
and oriented as indicated in the inset to Fig.~\ref{f1}. If the sail is assumed to be very thin, gas molecules hitting the sail will mostly transfer momentum perpendicular to PM, and in a much lesser extent parallel to it.
This corresponds in fact to a situation in which $\lambda_\parallel\ll\lambda_\perp$.

\begin{figure}[h]
\includegraphics[width=9cm,clip=true]{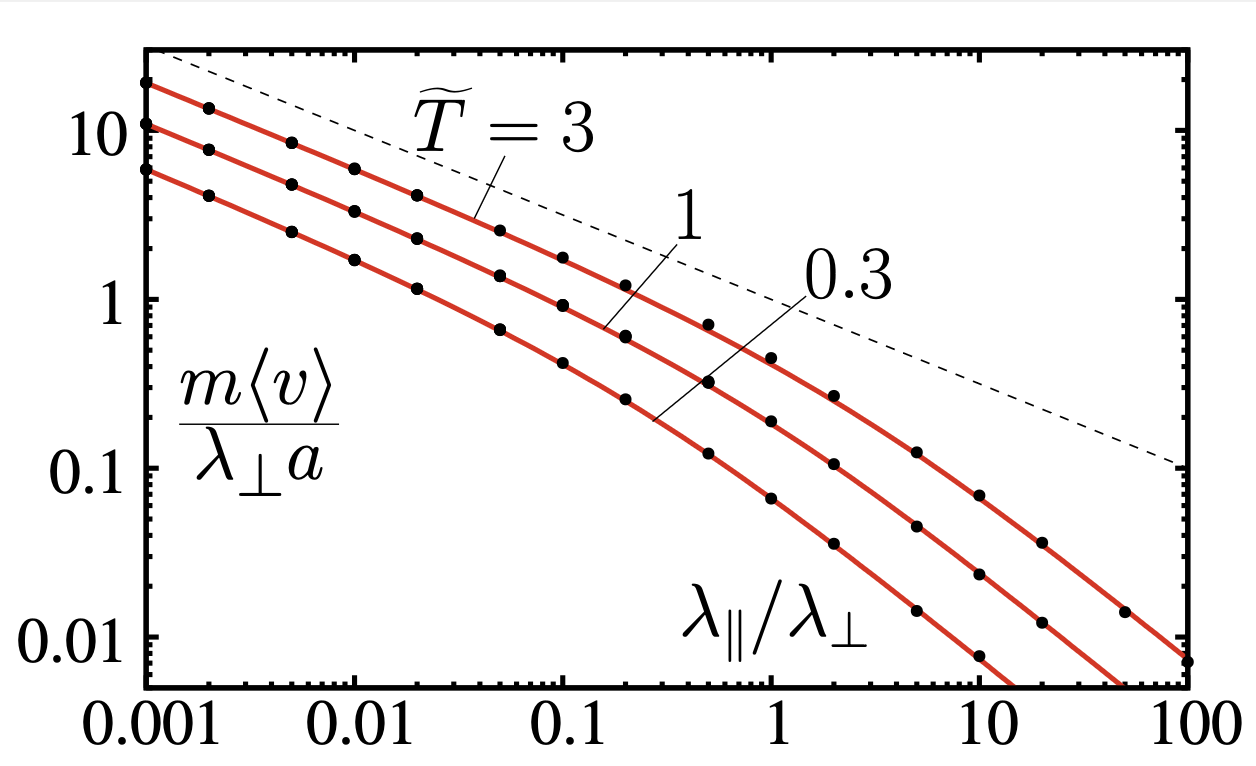}
\caption{Average velocity in numerical solutions of Eqs.~(\ref{2l}) and (\ref{2f}) (points), and analytical solution [Eq.~(\ref{vanal})] (continuous line). Parameters are $I/ma^2=1$, $\widetilde T\equiv mk_BT/a^2\lambda_\perp^2$ as indicated. We see how the typical velocity increases as $\lambda_\parallel/\lambda_\perp$ is reduced, and diverges as $(\lambda_\parallel/\lambda_\perp)^{-1}$ (dotted line has slope $-1$).
\label{two_coup}
}

\label{otramas}
\end{figure}

Eqs.~(\ref{2l}) and (\ref{2f}) can be treated analytically to obtain the long-time limiting behavior of $\langle v\rangle$. This treatment
is given in Appendix~\ref{AppB}. If $\lambda_\parallel\ne 0$, $\langle v\rangle$ tends to a well-defined finite limit as $t\to\infty$. This limiting value is shown in Fig.~\ref{otramas}, where we see that the limiting value of $\langle v\rangle$ becomes larger
as $\lambda_\parallel$ decreases, indicating that the translational kinetic energy $E=m \langle v\rangle^2/2$ may become much larger than the thermal energy $k_BT$, in apparent violation of the second law of thermodynamics.
To have an additional confirmation of this anomalous result, we performed numerical simulations of Eqs.~(\ref{2l}) and (\ref{2f}) (details of the numerics are contained in Appendix~\ref{AppC})
at different values of $\lambda_\parallel/\lambda_\perp$, and evaluated the average value of the velocity $\langle v \rangle$. The results (dots in Fig.~\ref{otramas})
nicely match the analytical results.

\begin{figure}[h]
\includegraphics[width=9cm,clip=true]{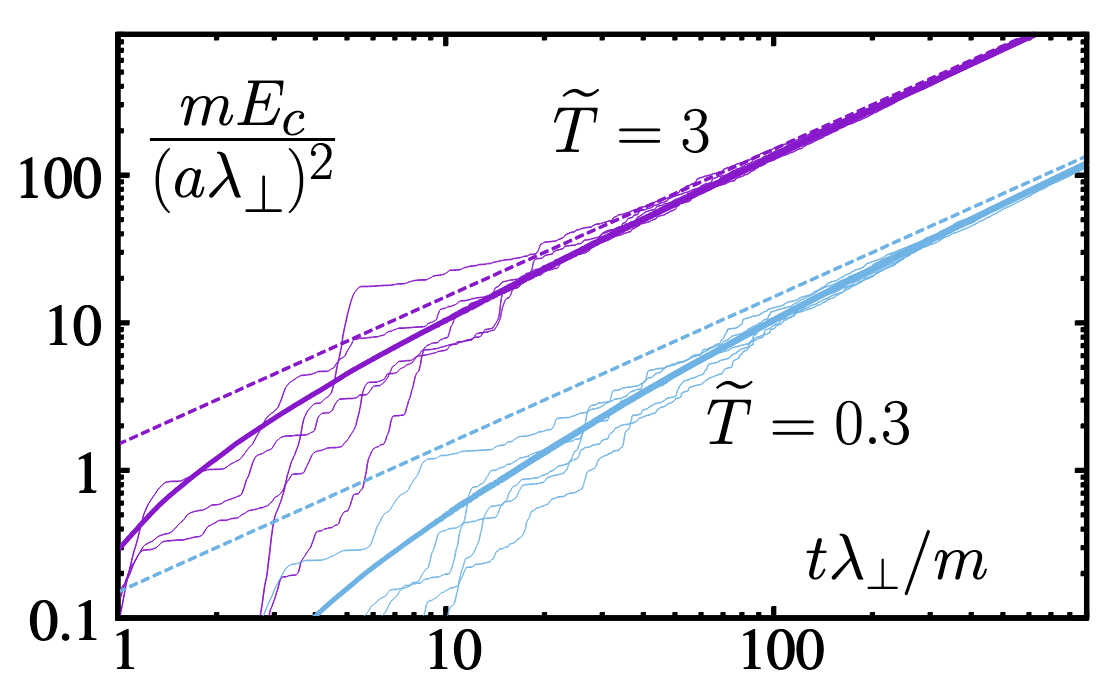}
\caption{Evolution of the translational kinetic energy $E_c\equiv mv^2/2$ from numerical simulations of Eqs.~(\ref{2l}) and (\ref{2f}),
with $\lambda_\parallel=0$ for two different values of the dimensionless temperature $\widetilde T\equiv mk_BT/a^2\lambda_\perp^2$, as indicated (see Appendix~\ref{AppC} for numerical details). The initial condition is $v=\omega=0$, and $I/ma^2=1$. Thin continuous lines are the results for five individual realizations, for each temperature. Thick continuous line is the result of averaging over 1000 realizations.
Dotted lines show the asymptotic analytic form from Eq.~(\ref{analytic}).
}

\label{f2}
\end{figure}

In order to have a simple explanation of this result, we will concentrate on the case $\lambda_\parallel=0$.
We will qualitatively analyze this unusual behavior in a limiting case in which the gas is rarefied in such a way that molecules hit the sail individually.
Let us assume we start with some initial condition $(v,\omega)\equiv(v_i,0)$. A random collision on the sail transfers momentum $\delta p$ perpendicular to PM in such a way that the new initial condition becomes
$\displaystyle \left(v_i,{\delta  p\over am}\right)$. For illustration we consider the case where the system completely transfers the rotation energy to the translation before the next collision. In this situation, the long-time condition that serves as the starting initial condition for the next step is
 $\displaystyle \left (\sqrt{v_i^2+\varepsilon} , 0 \right ) $, with $\varepsilon=(a^2+I/m) \left({\delta p}/{am}\right)^2$. In other words, after each collision the kinetic energy increases by $\Delta E= (a^2+I/m)\delta p^2/(2a^2 m)$. If $M$ is the mass of the gas particles, we obtain a linear increase in the energy of the form
\begin{equation}
    E = \left[ {M\over 2m} \left({1+{I\over ma^2}}\right)\right] \frac{k_BTt} {\tau},
    \label{Elinear}
\end{equation}
where $\tau $ is the average time between collisions of molecules with the sail and $T$ is the temperature of the gas.
This result can be verified analytically by integrating
Eqs.~(\ref{2l}) and (\ref{2f}) in the case $\lambda_\parallel=0$ (see Appendix~\ref{AppB}),
yielding (as in the qualitative previous argument) an average translational kinetic energy that increases linearly with time for long times:
\begin{equation}
\frac 12 m\langle v^2\rangle \simeq\frac{ a^2\lambda_\perp k_BT}{(a^2m+I)}t,
\label{analytic}
\end{equation}
coincident with our previous qualitative argument.
Figure~\ref{f2} shows a numerical simulation of Eqs.~(\ref{2l}) and (\ref{2f}) with $\lambda_\parallel=0$
confirming this intriguing behavior: if the NHC is taken at face value, coupling to a thermal bath only perpendicularly to the sleigh direction gives rise to a harvesting of energy from the thermal bath, in apparent violation of the second law of thermodynamics. The next section addresses this intriguing question, showing that the apparent breach of the second law
is a consequence of disregarding the influence of temperature within the constraint.
\cite{footnote1}

\section{The NHC as an infinite friction limit}

The violation of the second law of thermodynamics
is an indication that we have been inconsistent in the physical implementation of the constraint at finite temperature.
This brings us to a fundamental problem in the context of nonholonomic mechanics: can the NHC be obtained from a limiting behavior of a Lagrangian, holonomic system?
In general terms, this is still an open question.
However, NHCs can be regarded formally as originated in diverging friction forces. This interpretation goes
back at least to the work of
Caratheodory \cite{Caratheodory1933} who asked if the limiting case of such
friction could explain the motion of the Chaplygin sleigh.\cite{footnote2}
The general
case was considered by Karapetyan \cite{Karapetyan1983} and by Kozlov \cite{Kozlov1983} (see also \cite{Elderling2016}).
The key idea in the
nonholonomic setting is to
take a nonlinear Rayleigh dissipation function \cite{Bloch2003} proportional to an overall positive constant $\Lambda$, which plays the role of an effective friction coefficient.
Taking the limit as $\Lambda$ goes to infinity and using Tikhonov's theorem \cite{Tiko}
yields the nonholonomic dynamics.

The Chaplygin sleigh illustrates the case
in very nice fashion. Returning to Eqs.~(\ref{xy}) and (\ref{torque}), we will illustrate the limiting friction procedure by assuming (following Caratheodory and Fufaev) a friction condition in the lateral displacement of point P, namely $f= \Lambda u$. The equations now are written as ($\xi\equiv {I}/{a\Lambda}$)
\begin{subequations}
\begin{align}
u&=\xi\dot\omega, \label{fufaev_a}\\
\dot v&=\xi\omega\dot\omega+a\omega^2, \label{fufaev_b}\\
-\xi\ddot\omega&=v\omega+\frac{I+a^2m}{am}\dot\omega
\end{align}
\label{fufaev}
\end{subequations}
It is clear that, as the friction coefficient $\Lambda$ goes to infinity ($\xi\to 0$),
one recovers the classic
Eqs.~(\ref{clasico}).

The physical implementation of NHCs as limits of a viscous interaction provides a physically grounded resolution to the apparent violation of the second law of thermodynamics that we observed earlier.
However, if we intend to ``thermostatize'' the whole system to a single temperature bath at temperature $T$, the existence of a friction force with friction coefficient $\Lambda$ at point $P$ requires the presence of a stochastic Langevin force $\sim \sqrt{2\Lambda k_BT}\,\eta(t)$. Again, the amplitude of this term is fixed by the fluctuation-dissipation relation. The main effect of this term will be to produce a nonvanishing value of $u$ that, from the equipartition theorem can be estimated
to be $u\sim \sqrt {k_BT/m}$, independently of $\Lambda$. In other words, if
we strictly maintain the condition $u=0$, we are
considering the constraint to be at $T=0$. Therefore, if other parts of the
system are at non-zero $T$, we have a system operating between two different thermal baths at different temperatures, and extracting work from that system is not contradictory. This is in close analogy to the solution to the Feynman's ratchet problem \cite{FeynmanRatch}.

We now go back to Eq.~(\ref{nuevas}), and consider $f$ and $F_\perp$ to be the interaction forces with two different thermal baths
(we present the analysis for the case $F_\parallel=0$):
\begin{subequations}
\begin{align}
F_\perp&=\lambda_\perp U+\sqrt{2\lambda_\perp k_BT}\,\eta_\perp(t),         \\
f&=\Lambda u+\sqrt{2\Lambda k_BT_\Lambda}\,\eta_\Lambda(t),
\end{align}
\label{uf}
\end{subequations}
($U\equiv u+a\omega$).
The force $F_\perp$ is the coupling to the sail, as previously presented. The force $f$ is the friction-plus-stochastic force at P, intended to enforce the NHC as $\Lambda\to\infty$. Notice that for purposes of analysis we have allowed for the moment two different temperatures $T$, $T_\Lambda$.
The equations of motion can be written in this case as (see Appendix~\ref{AppA})
\begin{subequations}
\begin{align}
\dot v&=u\omega+a\omega^2, \label{uf2_a}\\
\dot u&=-\frac {F_\perp}m-v\omega-\frac{I+a^2m}{mI}f, \label{uf2_b}\\
\dot\omega&=\frac{af}{I} \label{uf2_c}
\end{align}
\label{uf2}
\end{subequations}

\begin{figure}[h]
\includegraphics[width=8cm,clip=true]{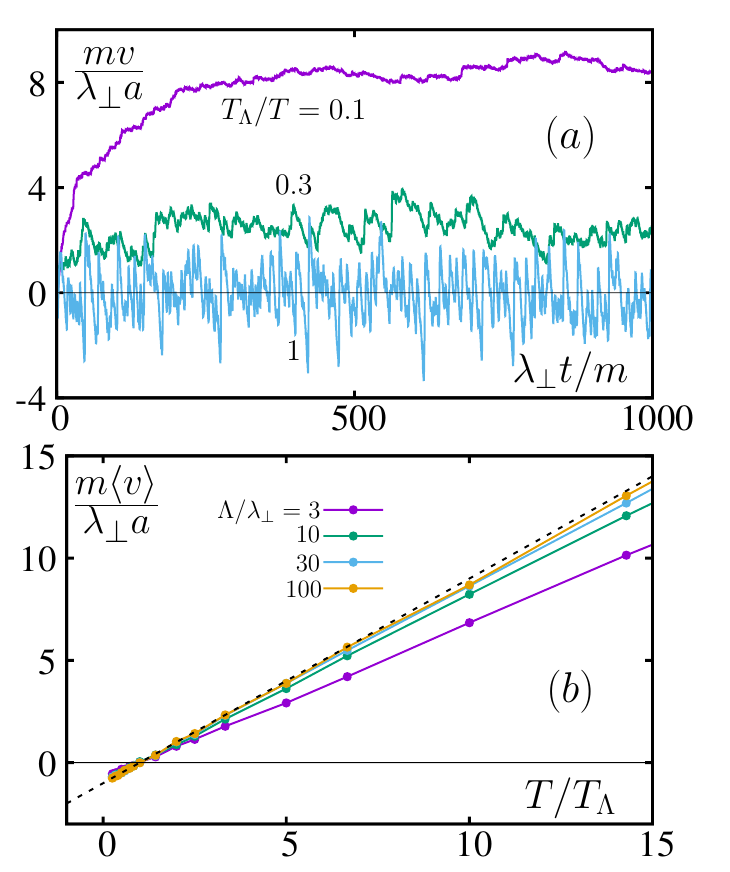}
\caption{(a) Time evolution of $v(t)$ from single realizations of Eqs.~(\ref{uf}) and (\ref{uf2}) (initial condition $u=v=\omega=0$), at different values of $T_\Lambda$, and with $\Lambda/\lambda_\perp=100$, $I/ma^2=1$, and $mk_BT/a^2\lambda_\perp^2=1$. (b) Asymptotic values of $\langle v\rangle$ as a function of $T/T_\Lambda$, and for different values of $\Lambda/\lambda_\perp$. Continuous straight line corresponds to $m\langle v\rangle/a\lambda_\perp=T/T_\Lambda-1$, and it is a guide to the eye.}
\label{f_chaply}
\end{figure}

Our main claim is that if the temperature at the constraint is considered to be the same as that of the bath interacting
with the sail (i.e.\ $T=T_\Lambda$), then thermodynamic consistency is restored, and no energy harvesting can occur, independently of the relative values of $\lambda_\perp$ and $\Lambda$. Numerical evidence is strong in this direction.
In fact, Fig.~\ref{f_chaply}(a)
presents results of numerical simulations of Eqs.~(\ref{uf}) and (\ref{uf2})
[see the traces $v(t)$ for cases with $T_\Lambda=T$, or $T_\Lambda<T$]. We see how the average value of $v$ sets asymptotically to zero as $T_\Lambda\to T$.
Fig.~\ref{f_chaply}(b) shows the asymptotic values of $\langle v\rangle$ when $T_\Lambda$ is varied, and for different values of $\Lambda/\lambda_\perp$, and we see again the $\langle v\rangle \to 0$ as $T_\Lambda\to T$, independently of the ratio $\Lambda/\lambda_\perp$. Notice also that in the case $T_\Lambda=0$, $T\ne 0$ there is an unbounded increase of $\langle v\rangle$ as $t\to\infty$, which is the result previously discussed in Sec.~II, for $\Lambda\to\infty$.

\section{Euler-Poincar\'e-Suslov equations}\label{EPSsection}

Similar behavior will occur in other systems, in particular
the so-called Euler-Poincar\'e-Suslov equations. Here we
briefly discuss these equations and another particular case, the Suslov problem, which
has essentially identical dynamics to the Chaplygin sleigh. More general
cases will be discussed in future work.

The general form for nonholonomic equations of motion may be found
in \cite{Bloch2003}. The equations arise from the Lagrange D'Alembert principle
and may be enforced by Lagrange multipliers or by writing the equations
in suitable variables that respect the constraints.

The general
equations of motion may be written as follows on $\mathbb{R}^n$: We are given
a Lagrangian $L(q,\dot{q})$ and nonholonomic constraints of the form
$\dot{s^a}+A(r,s)\dot{r^{\alpha}}=0$ where $q=(r,s)\in \mathbb{R}^{n-m}\times \mathbb{R}^m$.

One way to write the equations is in terms of the constrained Lagrangian $L_c$ which is formed
by substituting the constraints. ($L_c$ cannot be used as a regular Lagrangian
to give the dynamics. One has to use Lagrange D'Alembert.)

We obtain (see \cite{Bloch2003})
\[
\frac{d}{dt} \frac{\partial L_c}{\partial\dot r^\alpha}
- \frac{\partial L_c}{\partial r^\alpha} + A^a_\alpha \frac{\partial
L_c}{\partial s^a} = - \frac {\partial L}{\partial \dot s^b} B_{\alpha
\beta}^b \dot r^\beta
\]
where $B$ is a curvature term arising from the constraints.

If this had a cyclic variable, say $r^1$, then all the quantities
$L_c , L, B^b_{\alpha \beta}$ would be independent of $r^1$.
This is equivalent to saying that there is a translational symmetry
in the $r^1$ direction. Let us also suppose, as is often the case,
that the $s$ variables are also cyclic. Then the above equation for
the momentum $p_1 = \partial L_c / \partial \dot r^1$ becomes
\[
\frac{d}{dt}p_1  = - \frac{\partial L}{\partial
\dot s^b}  B_{1 \beta}^b \dot r^\beta.
\]
This fails to be a conservation law in general. Note that the
right-hand side is linear in $\dot{r}$ (the first term is linear in
$p_r$), and the equation does not depend on $r^1$ itself. So even the
presence of symmetry momentum is not always conserved.

We are interested in the situation where the configuration space is
$Q = G $, a Lie group.

In this case the basic equations are the Euler--Poincar\'e equations
\begin{equation}
\frac{d}{dt} p_b = C^c_{ab} I^{ad} p_c p_d=C^c_{ab}p_c\omega^a ,
\label{EPS}
\end{equation}
where $p _a =I_{ab}\omega^b$, $\omega\in\mathfrak{g}$,
$p\in\mathfrak{g}^*$ to which we append the left-invariant constraint
\begin{equation}
\langle\mathbf{a},\omega\rangle =a_i\omega^i=0.
\label{liconstraint}
\end{equation}

Here $\mathbf{a} =a_ie^i\in\mathfrak{g}^*$ and
$\omega=\omega^ie_i$ where $e_i,\,i=1, \dots, n$ is a basis for
$\mathfrak{g}$ and $e^i$ its dual basis.

Using the Lagrange multiplier approach we have the equations
(Euler--Poincar\'e--Suslov)
\begin{equation}
\frac{d}{dt} p_b = C^c_{ab} I^{ad} p_c p_d+\mu
a_b=C^c_{ab}p_c\omega^a+
\mu a_b
\label{EPScon}
\end{equation}
together with the constraint (\ref{liconstraint}). This defines a system
on the hyperplane defined by the constraints.

For the Chaplygin sleigh above, our equations are in the special euclidean group.
Below we consider the orthogonal group. We shall consider the general situation
in a forthcoming paper.

\subsection{The Suslov Problem}

The Suslov problem is a particular case of the Euler-Poincar\'e-Suslov equations. It describes the movement of a rigid body with angular velocity $\omega=(\omega_1,\omega_2,\omega_3)$, in a frame where the inertia matrix is of the form $I={\rm
diag}(I_1,I_2,I_3)$. The body is subject to the constraint
\begin{equation}
a\cdot \omega=0,
\label{Eulercon}
\end{equation}
where $a=(a_1,a_2,a_3)$.

The nonholonomic equations of motion are then given by
\begin{equation}
I\dot{\omega}=I\omega\times\omega+\mu a
\label{Suslov}
\end{equation}
subject to the constraint (\ref{Eulercon}). We can easily solve for
$\mu$:
\begin{equation}
\mu=-\frac{I^{-1} a \cdot(I\omega\times\omega)}
{I^{-1} a\cdot a}.
\label{mu_eq}
\end{equation}

If $a_2=a_3=0$ (a constraint that is an eigenstate of the moment
of inertia operator), one gets evolution with constant angular velocity.
This is the only situation when the classical Suslov
problem is measure preserving.

Now suppose $\omega_0$ is an equilibrium value. Taking the scalar product
of (\ref{Suslov}) with $I\omega_0$ yields
\begin{equation}
    a\cdot I\omega_0=Ia\cdot\omega_0=0.
    \label{eq_aIomega}
\end{equation}
Then, since $a$ is perpendicular to both $\omega_0$ and $I\omega_0$ we must have
\begin{equation}
    \omega_0=\beta Ia\times a
    \label{eq_omega0}
\end{equation}
for some $\beta$. The absolute value of $\beta$ can be determined from energy conservation from the initial condition. It can be shown that the asymptotic state corresponds to negative $\beta$ \cite{Kozlov2015}.

The Suslov problem for $a$ not an eigenvector of $I$ has the same phase portrait as
the Chapylgin sleigh shown in Fig.~\ref{Phase} (see \cite{Kozlov2015}) and thus the same apparent violation of the second law when coupled to a thermal bath.

\section{Conclusions}

In conclusion, we have shown that nonholonomic constraints must be treated with particular care at finite temperatures.
When such a constraint is naively enforced in the equations of motion
a striking outcome emerges: the second law of thermodynamics is violated. We have demonstrated this in the specific case of the Chaplygin sleigh, but the underlying mechanism is general and applies to systems where the nonholonomic dynamics, while conserving energy, fail to preserve phase-space volume. Through simple analytical arguments and full numerical integration of the equations of motion, we show that the constraint permits energy to be systematically extracted from the thermal bath. As a result, instead of relaxing toward thermal equilibrium, the system's energy grows without bound. We emphasize that the naive implementation of such a constraint at the level of the equations of motion tacitly assumes that the constraint itself is at zero temperature.
We uncovered this effect by analyzing in detail the implementation of nonholonomic constraints as the limiting case of friction forces with diverging viscosity. Realizing such constraints in physical systems, in a way that aligns with experimental observations, remains an open and intriguing challenge. Our results impose fundamental limits on these realizations: any implementation that rigidly enforces the constraint at the dynamic level, without additional compensating effects, is physically untenable.

{\bf Acknowledgments:} A.B. is supported in part by NSF Grant No.\ DMS-2103026 and AFOSR Grants No.\ FA9550-23-1-0215 and No.\ FA9550-23-1-0400.

\section*{Data Availability}
The data that support the findings of this article are openly available \cite{dataverse}, embargo periods may apply.

\appendix

\section{Appendix A}\label{AppA}

Here we give the details on how to pass from the equations of motion in Newtonian form to the canonical form, in term of $(v, \omega)$ variables.
We take one particular case, which is that of Eq.~(\ref{nuevas}) (the other possible cases are similar), reproduced here to facilitate the analysis:
\begin{subequations}
\begin{align}
m\ddot X&=(f+F_\perp)\sin(\theta), \label{A1a}\\
m\ddot Y&=-(f+F_\perp)\cos(\theta), \label{A1b}\\
I\ddot \theta&= af, \label{A1c}
\end{align}
\end{subequations}
where forces $f$ and $F_\perp$ are applied perpendicular to PM at points $P$ and M, respectively. The values of velocities $u$, $U$ and $v$ (Fig.~\ref{f1}) can thus be written as
\begin{subequations}
\begin{align}
v&=\dot X \cos\theta+\dot Y\sin\theta, \label{A2a}\\
U&=-\dot X \sin\theta+\dot Y \cos\theta, \label{A2b}\\
u&=U-a\dot\theta
\end{align}
\end{subequations}
Taking a time derivative and using (\ref{A1c}) we can readily verify that ($\omega\equiv\dot\theta$)
\begin{subequations}
\begin{align}
\dot v&=u\omega+a\omega^2, \label{A3a}\\
\dot u&=-\frac{f+F_\perp}m-\omega v -a\dot\omega, \label{A3b}\\
I\dot \omega&=af \label{A3c}
\end{align}
\end{subequations}
These are equivalent to Eq.~(\ref{uf2}) of Sec.~III.
From here, Eq.~(\ref{2l}) (Sec.~II\,B) are obtained by forcing $u=0$. If in addition we set $F_\parallel=0$, we obtain Eq.~(\ref{clasico})
(Sec.~II\,A). Finally, using $f\equiv Iu/a\xi$ we recover the Fufaev-Charatheodory Eqs.~(\ref{fufaev}).

\section{Appendix B}\label{AppB}

Here we show how the asymptotic behavior of the velocity [Eq.~(\ref{analytic})] is obtained from the
Chaplygin equations [Eqs.~(\ref{2l}) and (\ref{2f})] in the presence of a thermal bath.
For convenience, we reproduce the starting equations here
\begin{subequations}
\begin{align}
\dot v&=a\omega^2-\frac{F_\parallel}m, \label{B1a}\\
\dot\omega&=-\frac{am}{I+a^2m}  \left ({v\omega+\frac {F_\perp}m}\right ) \label{B1b}
\end{align}
\end{subequations}
with the forces
\begin{subequations}
\begin{align}
F_\parallel&=-\lambda_\parallel v +\sqrt{2\lambda_\parallel k_BT }\,\eta_\parallel(t), \label{B2a}\\
F_\perp&=-\lambda_\perp \omega a +\sqrt{2\lambda_\perp k_BT }\,\eta_\perp(t). \label{B2b}
\end{align}
\end{subequations}
Substituting (\ref{B2b}) into (\ref{B1b}) yields
\begin{equation}
\dot\omega=-\frac{a}{a^2m+I}\left (\left (mv+\lambda_\perp a\right )\omega+  {\sqrt{2\lambda_\perp k_B T}}\,\eta_\perp(t)\right ). \label{B3}
\end{equation}
Formally integrating this equation we obtain
\begin{equation}
\omega(t)=\frac{a\sqrt{2\lambda_\perp k_BT}}{a^2m+I}\int_{-\infty}^t dt'\,e^{-(t-t')/\tau}\eta_\perp(t')
\label{B4}
\end{equation}
with
\begin{equation}
\tau=\frac{a^2m+I}{a(mv+\lambda_\perp a)}.
\label{B5}
\end{equation}
In principle, $\tau$ is a time-dependent quantity through its $v$-dependence.
In order to proceed further we will assume $v(t)$ does not vary much during the time interval contributing to the integral in Eq.~(\ref{B4}). This is fulfilled if $v$ does not vary much during a time $\tau_0\equiv(a^2m+I)/a^2\lambda_\perp$.
Thus, considering $\tau$ as a constant in Eq.~(\ref{B4}),
we calculate $\langle \omega ^2\rangle $ as
\begin{equation}
\langle \omega ^2\rangle=\frac{2a^2 \lambda_\perp k_BT}{(a^2m+I)^2}\int_{-\infty}^t dt'\int_{-\infty}^t dt''\,  e^{-(2t-t'-t'')/\tau} \langle \eta_\perp(t')\eta_\perp(t'')\rangle
\label{B6}
\end{equation}
and using
\begin{equation}
\langle \eta_\perp(t')\eta_\perp(t'')\rangle=\delta(t'-t'')
\label{B7}
\end{equation}
we obtain
\begin{eqnarray}
\langle \omega ^2\rangle&=&\frac{2a^2 \lambda_\perp k_BT}{(a^2m+I)^2}\int_{-\infty}^t dt'\, e^{-2(t-t')/\tau} \label{B8}\\
&=&\frac{a^2 \lambda_\perp k_BT\tau}{(a^2m+I)^2} \label{B9}\\
&=&\frac{a \lambda_\perp k_BT}{(a^2m+I)(mv+\lambda_\perp a)}. \label{B10}
\end{eqnarray}
Now Eqs.~(\ref{B1a}) and (\ref{B2a}) give
\begin{equation}
\frac{d \langle v\rangle}{dt}=\frac{ a^2\lambda_\perp k_BT}{(a^2m+I)(m\langle v\rangle +\lambda_\perp a)}-\lambda_{\parallel}\langle v\rangle.
\label{B11}
\end{equation}
If $\lambda_\parallel\ne 0$, this equation provides a finite asymptotic value
\begin{equation}
\langle v \rangle (t\to\infty)=\frac{\lambda_\perp a}{2m} \left (\sqrt{1+\frac{4mk_BT}{(a^2m+I)a\lambda_\perp\lambda_\parallel}}-1\right ).
\label{vanal}
\end{equation}
This is the result plotted in Fig.~\ref{otramas}.
If $\lambda_\parallel= 0$, Eq.~(\ref{B11}) can be integrated to obtain
\begin{equation}
\frac 12 m\langle v^2\rangle+\lambda_\perp a \langle v\rangle=\frac{ a^2\lambda_\perp k_BT}{a^2m+I}t
\label{B13}
\end{equation}
(we are assuming $v(t=0)=0$). This expression gives a diverging value of $\langle v\rangle$ as $t\to\infty$. If we are interested in the asymptotic limit
the linear term can be neglected and we obtain
\begin{equation}
\frac 12 m\langle v^2\rangle =\frac{ a^2\lambda_\perp k_BT}{a^2m+I}t,
\label{B14}
\end{equation}
i.e., we obtain Eq.~(\ref{analytic}).

\section{Appendix C}\label{AppC}

Here we provide some details on the numerical simulations performed.

The kind of equation we numerically solve can all be written in the form\footnote{Equation~(\ref{uf2_b}) actually has two different stochastic terms on the right-hand side. The generalization to allow for this case is straightforward, and we avoid it here for simplicity of presentation.}
\begin{equation}
\dot {\overline x}={\overline f}({\overline x})+{\overline \eta}(t)
\label{C1}
\end{equation}
where the vector ${\overline x}$ stands for the pair $(v,\omega)$ for the results of Fig.~\ref{otramas}, or the triplet $(v,\omega,u)$ for those of Fig.~\ref{f2}. The stochastic terms ${\overline \eta}$ have zero mean and some fixed variances appropriate for each case:
\begin{equation}
\langle \eta_i(t)\eta_j(t')\rangle=\sigma^2_i\delta(t-t')\delta_{i,j}.
\label{C2}
\end{equation}
Our numerical scheme is a simple forward in time Euler method, where
\begin{equation}
{\overline x}^{(n+1)}={\overline x}^{(n)}+\left ({\overline f}\left ({\overline x}^{(n)}\right )+\overline \eta^{(n)}\right ) \Delta t
\label{C3}
\end{equation}
with superscripts denoting temporal steps, and $\overline \eta^{(n)}$ being discrete stochastic variables such that
\begin{equation}
\langle \eta^{(n)}_i(t)\eta^{(n')}_j(t')\rangle=(\Delta t)^{-1}\sigma^2_i\delta_{n,n'}\delta_{i,j}.
\label{C4}
\end{equation}
Despite its drawbacks, we prefer this simple first-order Euler scheme compared to more sophisticated methods (i.e., symplectic, and their generalizations to stochastic systems \cite{ngj1,ngj2}), because of its much simpler implementation. In any case we will show now that for our needs, the present approach provides results that nicely converge to a well-defined limit as $\Delta t$ is reduced.

In particular, we show here an example for Eqs.~(\ref{2l}) and (\ref{2f}), which avoids nonessential parameters and for $F_\parallel=0$
can be written as:
\begin{subequations}
\begin{align}
\dot v&=\omega^2, \label{C5a}\\
\dot\omega&=-\omega v-F_\perp(t), \label{C5b}\\
F_\perp(t)&=-\lambda_\perp \omega +\sqrt{2\lambda_\perp T}\,\eta_\perp(t)
\end{align}
\label{dless}
\end{subequations}

\begin{figure}[h]
\includegraphics[width=8cm,clip=true]{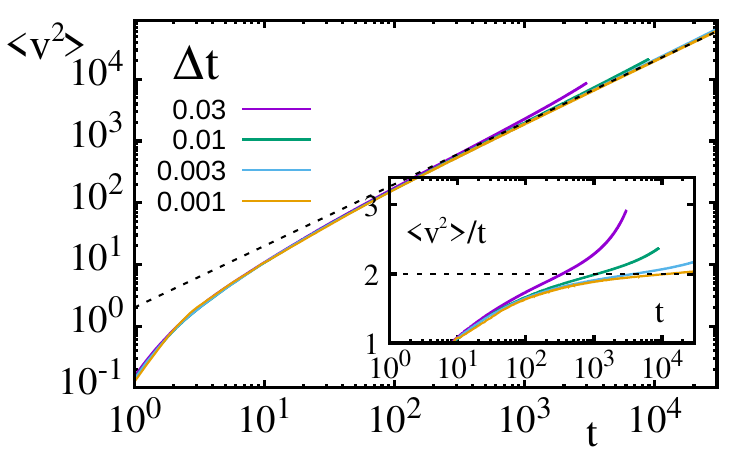}
\caption{Numerical solution to Eq.~(\ref{dless}) for different values of $\Delta t$, with $T\equiv 1$, $\lambda_\perp\equiv1$. An average over 100 realizations is shown.
Continuous straight lines show the expected analytic limiting behavior at large time, namely $\langle v^2\rangle=2t$.
}
\label{fig_apendice}
\end{figure}

We show in Fig.~\ref{fig_apendice} the results of numerical integration (for $T\equiv 1$, $\lambda_\perp \equiv 1$) and the obtained value for $\langle v^2\rangle$ (with averaging over 100 realizations), using different values of $\Delta t$. Our interest is in the asymptotic regime of large $t$ ($t\gg 1$) in which $\langle v^2\rangle/2\simeq t$. The results in Fig.~\ref{fig_apendice} corroborate this behavior, although there are deviations (and actually a divergence of the integration algorithm) at very large $t$. However, we see how the range in which a linear-in-$t$ behavior is observed increases as $\Delta t$ is reduced. In all simulations presented in the paper we have used $\Delta t=10^{-3}$.

\begin{thebibliography}{99}
\bibitem{footnote1}
Note that a similar question emerged in the classical Feynman's treatment of a thermal ratchet \cite{FeynmanRatch}, which in a superficial analysis gives rise to a violation of the second law.
\end{thebibliography}
\end{document}